\documentclass[12pt]{article}
\usepackage{graphicx}

\begin{document}

\title{\textbf{Influence of entrance channels on formation of superheavy nuclei in massive fusion reactions}}

\author{Zhao-Qing Feng$^{a}$\footnote{Corresponding author. Tel. +86 931 4969215. \newline \emph{E-mail address:} fengzhq@impcas.ac.cn},
Jun-Qing Li$^{a}$, Gen-Ming Jin$^{a}$}
\date{}
\maketitle

\begin{center}
$^{a}${\small \emph{Institute of Modern Physics, Chinese Academy of
Sciences, Lanzhou 730000, China}}\\[0pt]
\end{center}

\textbf{Abstract}
\par
Within the framework of the dinuclear system (DNS) model, the
production cross sections of superheavy nuclei Hs (Z=108) and Z=112
combined with different reaction systems are analyzed
systematically. It is found that the mass asymmetries and the
reaction Q values of the combinations play a very important role on
the formation cross sections of the evaporation residues. Both
methods by solving the master equations along the mass asymmetry
degree of freedom (1D) and along the proton and the neutron degrees
of freedom (2D) are compared each other and with the available
experimental results.

The synthesis of heavy or superheavy nuclei (SHN) is a very
important subject in nuclear physics motivated with respect to the
island of stability which is predicted theoretically, and has
obtained much experimental research with fusion-evaporation
reactions $\cite{Ho00,Og07}$. Combinations with a doubly magic
nucleus or nearly magic nucleus are usually chosen owing to the
larger reaction $Q$ values. Six new elements with Z=107-112 were
synthesized in cold fusion reactions for the first time and
investigated at GSI (Darmstadt, Germany) with the heavy-ion
accelerator UNILAC and the SHIP separator $\cite{Ho00,Mu99}$.
Recently, experiments on the synthesis of element 113 in the
$^{70}$Zn+$^{209}$Bi reaction have been performed successfully at
RIKEN (Tokyo, Japan) $\cite{Mo04}$. However, it is difficulty to
produce heavier SHN in the cold fusion reactions because of the
smaller production cross sections that are lower than 1 pb for
$Z>113$. The superheavy elements Z=113-116, 118 were synthesized at
FLNR in Dubna (Russia) with the double magic nucleus $^{48}$Ca
bombarding actinide nuclei $\cite{Og99,Og04a,Og06}$. New heavy
isotopes $^{259}$Db and $^{265}$Bh have also been synthesized at
HIRFL in Lanzhou (China) $\cite{Ga01}$. Further experimental works
are necessary in order to testify the new synthesized SHN. A better
understanding of the formation of SHN in the massive fusion
reactions is still a challenge for theory.

In this letter, we focus on the influence of the entrance mass
asymmetry and the reaction Q value of projectile-target combinations
on the production cross sections of superheavy residues. In the DNS
model, the evaporation residue cross section is expressed as a sum
over partial waves with angular momentum $J$ at the centre-of-mass
energy $E_{c.m.}$ $\cite{Fe06,Fe07,Fe09}$,
\begin{equation}
\sigma_{ER}(E_{c.m.})=\frac{\pi \hbar^{2}}{2\mu
E_{c.m.}}\sum_{J=0}^{J_{max}}(2J+1)
T(E_{c.m.},J)P_{CN}(E_{c.m.},J)W_{sur}(E_{c.m.},J).
\end{equation}
Here, $T(E_{c.m.},J)$ is the transmission probability of the two
colliding nuclei overcoming the Coulomb potential barrier in the
entrance channel to form the DNS. The $P_{CN}$ is the probability
that the system will evolve from a touching configuration into the
compound nucleus in competition with quasi-fission of the DNS and
fission of the heavy fragment. The last term is the survival
probability of the formed compound nucleus, which can be estimated
with the statistical evaporation model by considering the
competition between neutron evaporation and fission $\cite{Fe06}$.
We take the maximal angular momentum as $J_{max}=30$ since the
fission barrier of the heavy nucleus disappears at high spin
$\cite{Re00}$.

In order to describe the fusion dynamics as a diffusion process
along proton and neutron degrees of freedom, the fusion probability
is obtained by solving a set of master equations numerically in the
potential energy surface of the DNS. The time evolution of the
distribution probability function $P(Z_{1},N_{1},E_{1},t)$ for
fragment 1 with proton number $Z_{1}$ and neutron number $N_{1}$
with excitation energy $E_{1}$ is described by the following master
equations $\cite{Hu08}$,
\begin{eqnarray}
\frac{d P(Z_{1},N_{1},E_{1},t)}{dt}=\sum_{Z_{1}^{\prime
}}W_{Z_{1},N_{1};Z_{1}^{\prime},N_{1}}(t)\left[
d_{Z_{1},N_{1}}P(Z_{1}^{\prime},N_{1},E_{1}^{\prime},t)-d_{Z_{1}^{\prime
},N_{1}}P(Z_{1},N_{1},E_{1},t)\right]+
\nonumber \\
\sum_{N_{1}^{\prime }}W_{Z_{1},N_{1};Z_{1},N_{1}^{\prime}}(t)\left[
d_{Z_{1},N_{1}}P(Z_{1},N_{1}^{\prime},E_{1}^{\prime},t)-d_{Z_{1},N_{1}^{\prime}}P(Z_{1},N_{1},E_{1},t)\right]-
\nonumber \\
\left[\Lambda^{qf}(\Theta(t))+\Lambda^{fis}(\Theta(t))
\right]P(Z_{1},N_{1},E_{1},t).
\end{eqnarray}
Here $W_{Z_{1},N_{1};Z_{1}^{\prime},N_{1}}$
($W_{Z_{1},N_{1};Z_{1},N_{1}^{\prime}}$) is the mean transition
probability from the channel $(Z_{1},N_{1},E_{1})$ to
$(Z_{1}^{\prime},N_{1},E_{1}^{\prime})$ (or $(Z_{1},N_{1},E_{1})$ to
$(Z_{1},N_{1}^{\prime},E_{1}^{\prime})$) , and $d_{Z_{1},N_{1}}$
denotes the microscopic dimension corresponding to the macroscopic
state $(Z_{1},N_{1},E_{1})$. The sum is taken over all possible
proton and neutron numbers that fragment
$Z_{1}^{\prime},N_{1}^{\prime}$ may take, but only one nucleon
transfer is considered in the model with $Z_{1}^{\prime }=Z_{1}\pm
1$ and $N_{1}^{\prime }=N_{1}\pm 1$. The excitation energy $E_{1}$
is determined by the dissipation energy from the relative motion and
the potential energy surface of the DNS. The motion of nucleons in
the interacting potential is governed by the single-particle
Hamiltonian $\cite{Fe06,Fe07}$. The evolution of the DNS along the
variable R leads to the quasi-fission of the DNS. The quasi-fission
rate $\Lambda^{qf}$ and the fission rate $\Lambda^{fis}$ can be
estimated with the one-dimensional Kramers formula
$\cite{Fe07,Fe09}$.

In the relaxation process of the relative motion, the DNS will be
excited by the dissipation of the relative kinetic energy. The local
excitation energy is determined by the excitation energy of the
composite system and the potential energy surface of the DNS. The
potential energy surface (PES) of the DNS is given by
\begin{eqnarray}
U(Z_{1},N_{1},Z_{2},N_{2};J,\textbf{R};\beta_{1},\beta_{2},\theta_{1},\theta_{2})=B(Z_{1},N_{1})+B(Z_{2},N_{2})-
\left[B(Z,N)+V^{CN}_{rot}(J)\right]+
\nonumber \\
V(Z_{1},N_{1},Z_{2},N_{2};J,\textbf{R};\beta_{1},\beta_{2},\theta_{1},\theta_{2})
\end{eqnarray}
with $Z_{1}+Z_{2}=Z$ and $N_{1}+N_{2}=N$. Here $B(Z_{i},N_{i})
(i=1,2)$ and $B(Z,N)$ are the negative binding energies of the
fragment $(Z_{i},N_{i})$ and the compound nucleus $(Z,N)$,
respectively, in which the shell and the pairing corrections are
included reasonably. The $V^{CN}_{rot}$ is the rotation energy of
the compound nucleus. The $\beta_{i}$ represent the quadrupole
deformations of the two fragments. The $\theta_{i}$ denote the
angles between the collision orientations and the symmetry axes of
deformed nuclei. The interaction potential between fragment
$(Z_{1},N_{1})$ and $(Z_{2},N_{2})$ includes the nuclear, Coulomb
and centrifugal parts, the details are given in Ref. $\cite{Fe07}$.
In the calculation, the distance $\textbf{R}$ between the centers of
the two fragments is chosen to be the value which gives the minimum
of the interaction potential, in which the DNS is considered to be
formed. So the PES depends on the proton and neutron numbers of the
fragment. In Fig.1 we give the potential energy surface in the
reaction $^{30}$Si+$^{252}$Cf as functions of the protons and
neutrons of the fragments in the left panel. The incident point is
shown by the solid circle and the minimum way in the PES is added by
the thick line. The driving potential as a function of the mass
asymmetry that was calculated in Ref. $\cite{Fe06,Fe07}$ is also
given in the right panel and compared with the minimum way in the
left panel. The driving potential at the incident point in 1D PES is
located at the maximum value, so there is no the inner fusion
barrier for the system, which results in a too large fusion
probability. Therefore, we solve the master equations within the 2D
PES to get the fusion probability for the systems with larger mass
asymmetries.

The formation probability of the compound nucleus at the Coulomb
barrier $B$ (here a barrier distribution $f(B)$ is considered) and
for angular momentum $J$ is given by$\cite{Fe06,Fe07}$
\begin{equation}
P_{CN}(E_{c.m.},J,B)=\sum_{Z_{1}=1}^{Z_{BG}}\sum_{N_{1}=1}^{N_{BG}}P(Z_{1},N_{1},E_{1},\tau
_{int}(E_{c.m.},J,B)).
\end{equation}
We obtain the fusion probability as
\begin{equation}
P_{CN}(E_{c.m.},J)=\int f(B)P_{CN}(E_{c.m.},J,B)dB,
\end{equation}
where the barrier distribution function is taken in asymmetric
Gaussian form.

The survival probability of the excited compound nucleus cooled by
the neutron evaporation in competition with fission is expressed as
follows:
\begin{equation}
W_{sur}(E_{CN}^{\ast},x,J)=P(E_{CN}^{\ast},x,J)\prod\limits_{i=1}^{x}\left(
\frac{\Gamma _{n}(E_{i}^{\ast},J)}{\Gamma
_{n}(E_{i}^{\ast},J)+\Gamma _{f}(E_{i}^{\ast},J)}\right) _{i},
\end{equation}
where the $E_{CN}^{\ast}, J$ are the excitation energy and the spin
of the compound nucleus, respectively. The $E_{i}^{\ast}$ is the
excitation energy before evaporating the $i$th neutron, which has
the relation
\begin{equation}
E_{i+1}^{\ast}=E_{i}^{\ast}-B_{i}^{n}-2T_{i},
\end{equation}
with the initial condition $E_{1}^{\ast}=E_{CN}^{\ast}$. The energy
$B_{i}^{n}$ is the separation energy of the $i$th neutron. The
nuclear temperature $T_{i}$ is given as
$E_{i}^{\ast}=aT_{i}^{2}-T_{i}$ with the level density parameter
$a$. $P(E_{CN}^{\ast},x,J)$ is the realization probability of
emitting $x$ neutrons. The widths of neutron evaporation and fission
are calculated using the statistical model. The details can be found
in Refs. $\cite{Fe06,Fe09}$.

With this procedure introduced above, we calculated the evaporation
residue excitation functions using the 1D and 2D master equations in
the reaction $^{48}$Ca+$^{238}$U as shown in Fig.2 represented by
dashed and solid lines, respectively, and compared them with the
experimental data performed in Dubna $\cite{Og04b}$ and at GSI
$\cite{Ho07}$. The GSI results show that the formation cross
sections in the 3n channel at the same excitation energy with 35 MeV
have a slight decrease, which are in a good agreement with our 1D
calculations. In the whole range, the 2D calculations give smaller
cross sections than 1D master equations owing to the decrease of the
fusion probability. For the considered system, the value of the PES
at the incident point is located at the line of the minimum way. So
the 1D master equations can give reasonable results. However, for
the systems with larger mass asymmetries and larger quadrupole
deformation parameters, e.g. $^{16}$O+$^{238}$U,
$^{22}$Ne+$^{244}$Pu, etc, the 1D master equations give too large
fusion probabilities.

The synthesis of heavy or superheavy nuclei through fusing two
stable nuclei is inhibited by the so-called quasi-fission process.
The entrance channel combinations of projectile and target will
influence the fusion dynamics. The suppression of the evaporation
residue cross sections for less fissile compound systems such as
$^{216}$Ra and $^{220}$Th when reactions are involved in projectiles
heavier than $^{12}$C and $^{16}$O was observed in Refs.
$\cite{Be01}$. The wider width of the mass distributions for the
fission-like fragments was also reported in Ref. $\cite{Th08}$. In
Fig.3 we calculated the transmission and fusion probabilities using
the 2D master equations for the reactions $^{34}$S+$^{238}$U,
$^{64}$Fe+$^{208}$Pb and $^{136}$Xe+$^{136}$Xe which lead to the
same compound nucleus $^{272}$Hs formation. The larger transmission
probabilities were found in the reactions $^{64}$Fe+$^{208}$Pb and
$^{136}$Xe+$^{136}$Xe owing to the larger Q values (absolute
values). Smaller mass asymmetries of the two systems result in a
decrease of the fusion probabilities. The evaporation residue
excitation functions in 1n-5n channels are shown in Fig.4. The
competition of the capture and the fusion process of the three
systems leads to different trends of the evaporation channels. The
3n and 4n channels in the reaction $^{34}$S+$^{238}$U, 1n and 2n
channels in the reaction $^{64}$Fe+$^{208}$Pb are favorable to
produce the isotopes $^{269,268}$Hs and $^{271,270}$Hs. Although the
system $^{136}$Xe+$^{136}$Xe consists of two magic nuclei, the
higher inner fusion barrier decreases the fusion probabilities and
enhances the quasi-fission rate of the DNS, hence leads to the
smaller cross sections of the Hs isotopes. The upper limit cross
sections for evaporation residues $\sigma_{(1-3)n}\leq$4 pb were
observed in a recent experiment $\cite{Og09}$, which are much lower
than the ones predicted by the fusion by diffusion model
$\cite{Sw04}$. In the DNS model, the larger mass asymmetry favors
the nucleon transfer from the light projectile to heavy target, and
therefore enhances the fusion probability of two colliding nuclei.

The superheavy element Z=112 was synthesized at GSI with the new
isotope $^{277}$112 in cold fusion reaction $^{70}$Zn+$^{208}$Pb
$\cite{Ho95}$ and also fabricated with more neutron-rich isotopes
$^{282,283}$112 in $^{48}$Ca induced reaction $^{48}$Ca+$^{238}$U.
We analyzed the combinations $^{30}$Si+$^{252}$Cf,
$^{36}$S+$^{250}$Cm, $^{40}$Ar+$^{244}$Pu and $^{48}$Ca+$^{238}$U
which lead to the production of new isotopes of the element Z=112
between the cold fusion reactions and the $^{48}$Ca induced
reactions as shown in Fig.5. The 2n, 3n and 4n channels in the
reaction $^{30}$Si+$^{252}$Cf, and the 4n channel in the reaction
$^{36}$S+$^{250}$Cm have larger cross section to produce new
isotopes due to the larger fusion probabilities of the two colliding
nuclei.

In summary, we systematically analyzed the entrance channel effects
of synthesizing SHN using the DNS model. The systems with larger
entrance mass asymmetry and larger reaction Q value can enhance the
capture and fusion probabilities of two colliding nuclei.
Calculations were carried out for the reactions $^{34}$S+$^{238}$U,
$^{64}$Fe+$^{208}$Pb and $^{136}$Xe+$^{136}$Xe which lead to the
same compound nucleus formation. The 2n, 3n and 4n channels in the
reaction $^{30}$Si+$^{252}$Cf, and the 4n channel in the reaction
$^{36}$S+$^{250}$Cm are favorable to synthesize new isotopes of the
element Z=112 at the stated excitation energies.

\textbf{Acknowledgements}

We would like to thank Prof. Werner Scheid for carefully reading the
manuscript. This work was supported by the National Natural Science
Foundation of China under Grant No. 10805061, the special foundation
of the president fellowship, the west doctoral project of Chinese
Academy of Sciences, and major state basic research development
program under Grant No. 2007CB815000.

\newpage
\begin{figure}
\begin{center}
{\includegraphics*[width=0.8\textwidth]{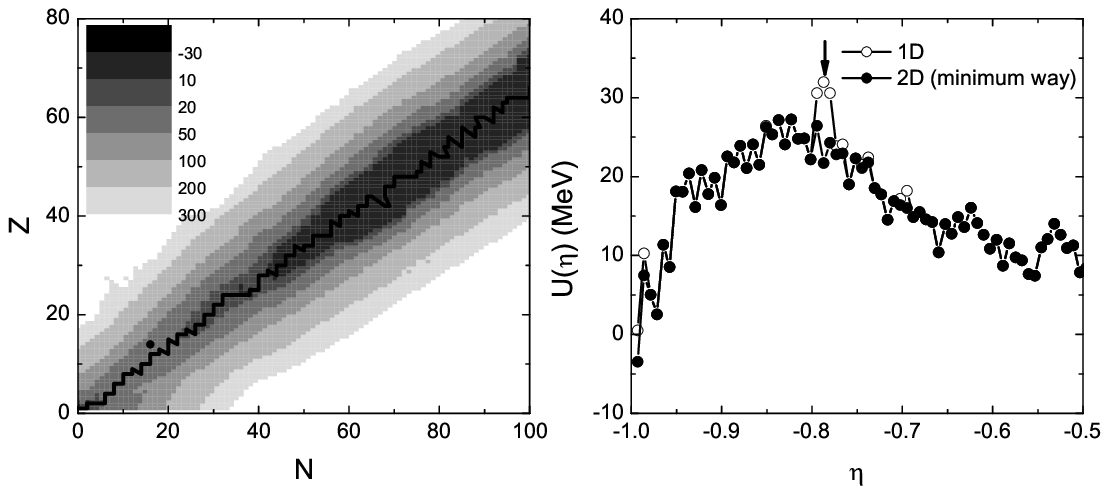}}
\end{center}
\caption{The potential energy surface of the DNS in the reaction
$^{30}$Si+$^{252}$Cf as functions of the protons and neutrons of the
fragments (left panel) and the mass asymmetry coordinate (right
panel).}
\end{figure}

\begin{figure}
\begin{center}
{\includegraphics*[width=0.6\textwidth]{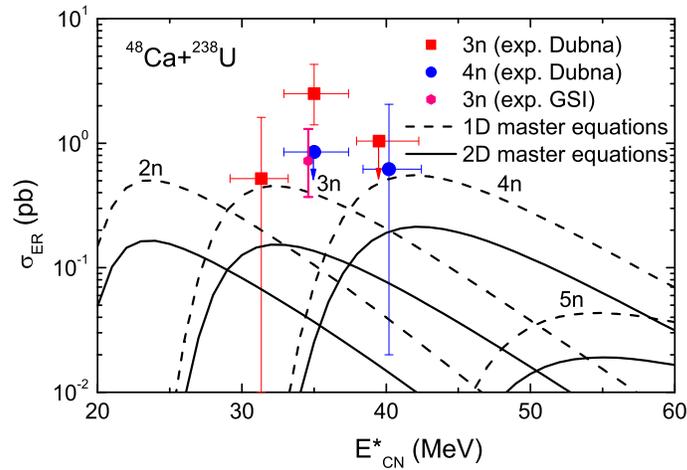}}
\end{center}
\caption{Comparison of the calculated evaporation residue excitation
functions using the 1D and 2D master equations with the available
experimental data in the reaction $^{48}$Ca+$^{238}$U.}
\end{figure}

\begin{figure}
\begin{center}
{\includegraphics*[width=0.8\textwidth]{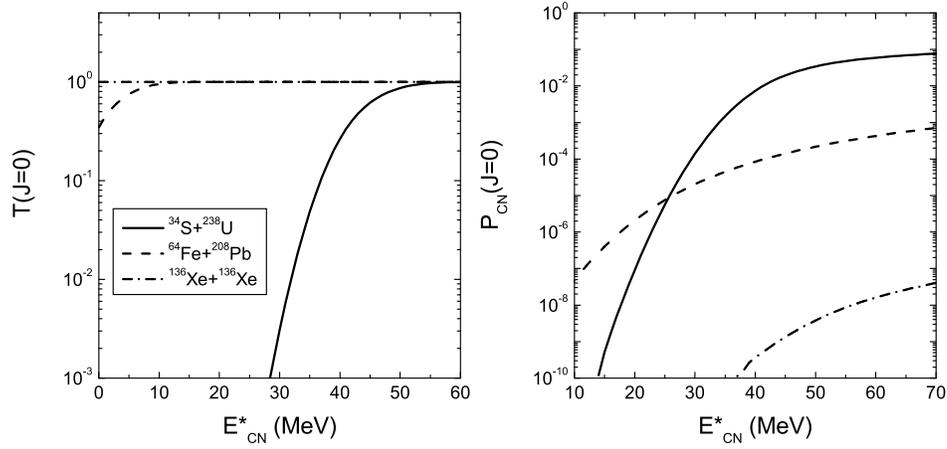}}
\end{center}
\caption{Calculated transmission and fusion probabilities as
functions of the excitation energies of the compound nucleus for the
reactions $^{34}$S+$^{238}$U, $^{64}$Fe+$^{208}$Pb and
$^{136}$Xe+$^{136}$Xe.}
\end{figure}

\begin{figure}
\begin{center}
{\includegraphics*[width=0.8\textwidth]{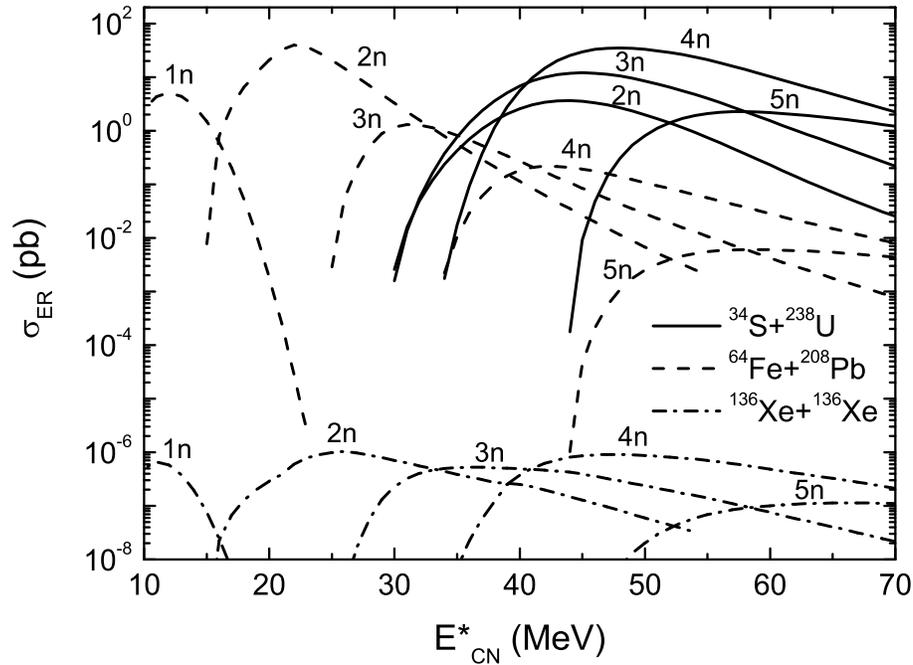}}
\end{center}
\caption{Comparison of the calculated evaporation residue cross
sections in 1n-5n channels using the 2D master equations for the
reactions $^{34}$S+$^{238}$U, $^{64}$Fe+$^{208}$Pb and
$^{136}$Xe+$^{136}$Xe.}
\end{figure}

\begin{figure}
\begin{center}
{\includegraphics*[width=0.8\textwidth]{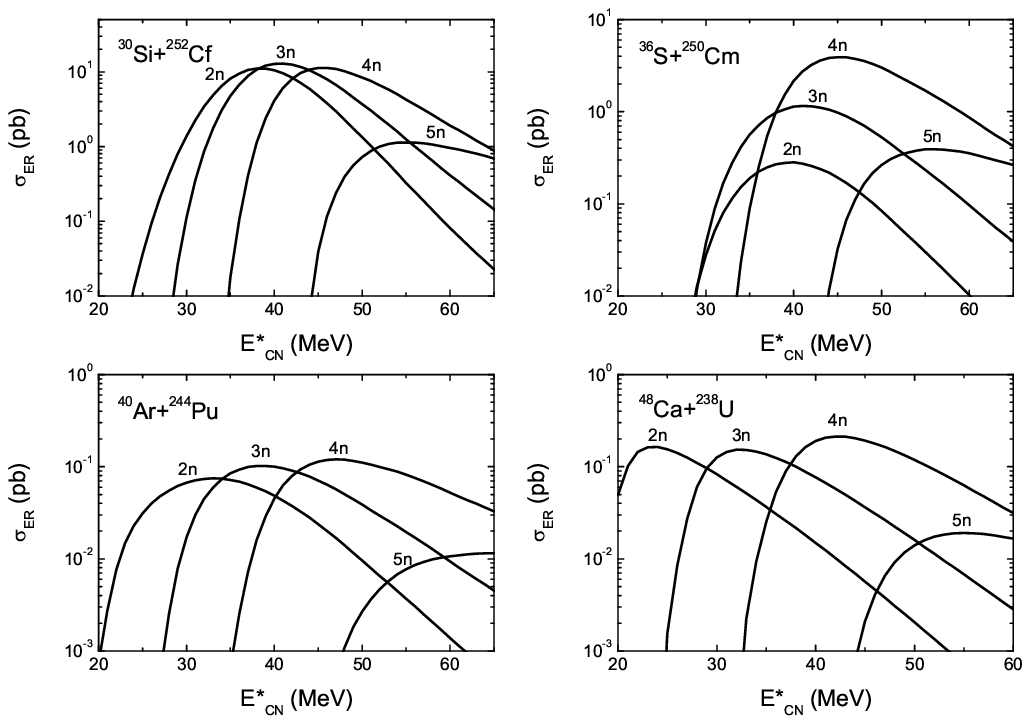}}
\end{center}
\caption{The same as in Fig.4, but for the reactions
$^{30}$Si+$^{252}$Cf, $^{36}$S+$^{250}$Cm, $^{40}$Ar+$^{244}$Pu and
$^{48}$Ca+$^{238}$U leading to the formation of the element Z=112.}
\end{figure}


\begin{thebibliography}{99}

\bibitem{Ho00} S. Hofmann, G. M\"{u}nzenberg, Rev. Mod. Phys. 72 (2000) 733;
S. Hofmann, Rep. Prog. Phys. 61 (1998) 639.
\bibitem{Og07} Yu.Ts. Oganessian, J. Phys. G 34 (2007) R165;
Nucl. Phys. A 787 (2007) 343c.
\bibitem{Mu99} G. M\"{u}nzenberg, J. Phys. G 25 (1999) 717.
\bibitem{Mo04} K. Morita, K. Morimoto, D. Kaji, et al., J. Phys. Soc.
Jpn. 73 (2004) 2593.
\bibitem{Og99} Yu.Ts. Oganessian, A.G. Demin, A.S. Iljnov, et al.,
Nature 400 (1999) 242; Yu.Ts. Oganessian, V.K. Utyonkov, Yu.V.
Lobanov, et al., Phys. Rev. C 62 (2000) 041604(R).
\bibitem{Og04a} Yu.Ts. Oganessian, V.K. Utyonkov, Yu.V. Lobanov, et al., Phys. Rev. C 69 (2004) 021601(R).
\bibitem{Og06} Yu.Ts. Oganessian, V.K. Utyonkov, Yu.V. Lobanov, et al., Phys. Rev. C 74 (2006) 044602.
\bibitem{Ga01} Z.G. Gan, Z. Qin, H.M. Fan, et al., Eur. Phys. J. A 10 (2001) 21;
Z.G. Gan, J.S. Guo, X.L. Wu, et al., Eur. Phys. J. A 20 (2004) 385.
\bibitem{Fe06} Z.Q. Feng, G.M. Jin, F. Fu, J.Q. Li, Nucl. Phys. A 771 (2006) 50.
\bibitem{Fe07} Z.Q. Feng, G.M. Jin, J.Q. Li, W. Scheid, Phys. Rev. C 76 (2007) 044606.
\bibitem{Fe09} Z.Q. Feng, G.M. Jin, J.Q. Li, W. Scheid, Nucl. Phys. A 816 (2009) 33.
\bibitem{Re00} P. Reiter, T.L. Khoo, T. Lauritsen, et al., Phys. Rev. Lett.
84 (2000) 3542.
\bibitem{Hu08} M.H. Huang, Z.G. Gan, Z.Q. Feng, et al., Chin. Phys. Lett. 25 (2008) 1243.
\bibitem{Og04b} Yu.Ts. Oganessian, V.K. Utyonkov, Yu.V. Lobanov, et al., Phys. Rev. C 70 (2004) 064609.
\bibitem{Ho07} S. Hofmann, D. Ackermann, S. Antalic, et al., Eur.
Phys. J. A 32 (2007) 251.
\bibitem{Be01} A.C. Berriman, D.J. Hinde, M. Dasgupta, et al.,
Nature 413 (2001) 144; D.J. Hinde, M. Dasgupta, A. Mukherjee, Phys.
Rev. Lett. 89 (2002) 282701.
\bibitem{Th08} R.G. Thomas, D.J. Hinde, D. Duniec, et al., Phys. Rev. C 77
(2008) 034610.
\bibitem{Og09} Yu.Ts. Oganessian, S.N. Dmitriev, A.V. Yeremin, et al., Phys. Rev. C 79 (2009) 024608.
\bibitem{Sw04} W.J. Swiatecki, K. Siwek-Wilczynska, J. Wilczynski,
Int. J. Mod. Phys. E 13 (2004) 261.
\bibitem{Ho95} S. Hofmann, V. Ninov, F.P. He{\ss}berger, et al., Z.
Phys. A 350 (1995) 277.

\end{thebibliography}
\end{document}